\documentstyle[12pt,epsf]{article}
\newcommand{\tdm}[1]{\mbox{\boldmath $#1$}}

\begin{document}
\begin{titlepage}
\pagestyle{empty}
\vspace*{1cm}
\begin{center}
{\large\bf  Diffractive $J/\Psi$ production in high energy $\gamma \gamma$ 
collisions as a probe of the QCD pomeron.}
\vspace{1.1cm}\\
         {\sc J.~Kwieci\'nski}$^a$,
         {\sc L.~Motyka\footnote{A fellow of the Foundation for Polish 
                                 Science}}$^b$,
\vspace{0.3cm}\\
$^a${\it Department of Theoretical Physics, \\
H.~Niewodnicza\'nski Institute of Nuclear Physics,
Cracow, Poland}
\vspace{0.3cm}\\
$^b${\it Institute of Physics, Jagellonian University,
Cracow, Poland}
\end{center}
\vspace{1.5cm}
\begin{abstract}
The reaction $\gamma \gamma \rightarrow J/\Psi J/\Psi$ is discussed 
assuming dominance of the QCD BFKL pomeron exchange.   We give prediction 
for the cross-section of this process for LEP2 
and TESLA energies.   We solve the BFKL equation in the non-forward 
configuration taking into account dominant non-leading 
effects which come from the requirement that the virtuality of the 
exchanged gluons along the 
gluon ladder is controlled by their transverse momentum squared.  
We compare our results with those corresponding to the simple two gluon 
exchange mechanism and with the BFKL pomeron exchange in the leading 
logarithmic 
approximation.  The BFKL effects are found to generate a steeper $t$-dependence 
than the two gluon exchange. The cross-section is found to  increase with 
increasing CM energy $W$ as   
$(W^2)^{2\lambda}$.  The parameter $\lambda$ 
is slowly 
varying with $W$ and takes the values $\lambda \sim 0.23 - 0.28$.   
  The magnitude of the total cross-section for the process 
$\gamma \gamma \rightarrow J/\Psi  J/\Psi$ 
is found to increase  from 4 to 26~pb  within 
the energy range accessible at  LEP2. 
The magnitude of the total cross-section for the process    
$e^+e^- \rightarrow e^+e^-J/\Psi  J/\Psi$ with antitagged  $e^+$ and $e^-$  
is estimated to be around 0.1 pb 
at LEP2. 
\end{abstract}
\vspace{1cm}

\noindent
%{\sf TPJU--14/98}\\
%{\sf June 1998}\\            
\end{titlepage} 

The high energy limit of elementary processes in perturbative QCD is
at present theoretically  fairly well understood
\cite{GLR,LIPAT1}.  The leading behaviour is controlled by   
the pomeron singularity which 
corresponds to the sum of ladder diagrams with reggeized 
gluons along the chain.  This sum  is described by
the Balitzkij, Fadin, Kuraev, Lipatov (BFKL) equation
\cite{BFKL1}.  Possible phenomenological tests of the
perturbative QCD pomeron exchange  are  however difficult.
First of all they have to be limited to (semi-) hard
processes in which the presence of the hard scale(s) can justify the
use of perturbative QCD. Moreover in order to minimize the possible
role of the non-perturbative contributions it is in principle necessary to
focus on the processes which directly probe the high energy
limit  of partonic amplitudes alone.  Finally in order to extract the genuine 
BFKL effects which go beyond the conventional QCD evolution with ordered 
transverse momenta from one scale to another it is also  
useful  to consider the processes with small (or equal to zero) 
"evolution length" i.e. those where the magnitudes of the two hard scales 
are comparable. The two  classical
processes which can probe the QCD pomeron by fulfilling these criteria 
are   deep inelastic events accompanied by an energetic (forward)
jet {\cite{MUELLERJ, DISJET}) and the production of large $p_T$
jets separated by the rapidity gap \cite{JETGAP}.  The former process 
probes the QCD pomeron in the forward direction while the latter
reflects the elastic scattering of partons via the QCD pomeron
exchange with  non-zero (and large) momentum transfer. Another possible probe of the QCD 
pomeron at (large) 
momentum transfers can be provided by the diffractive vector 
meson photoproduction accompanied by proton dissociation in order to avoid 
nucleon form-factor 
effects  \cite{FORSHAW,BARTLQ},  while the
complementary measurement to deep inelastic scattering + jet
events may also be the total $\gamma^* \gamma^*$ cross section 
of virtual photons having comparable virtuality \cite{GGSTAR}.\\

\noindent
\begin{figure}
\begin{center}
\leavevmode
\epsfxsize = 9cm
\epsfysize = 7cm 
%\epsfbox[50 565 355 815]{ppfig1.ps}
\epsfbox{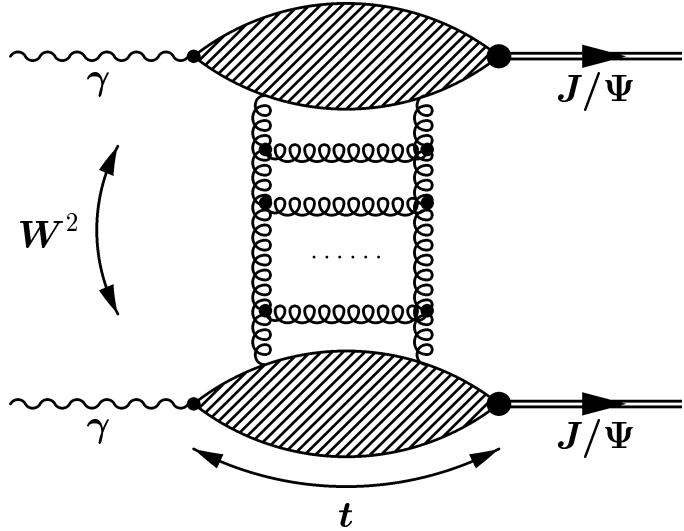}
\end{center}
\caption{
\small The QCD pomeron exchange mechanism of the process $\gamma
\gamma \rightarrow J/\Psi J/\Psi$.} 
\end{figure}

In this paper we wish to analyze the double diffractive production of
$J/\Psi$ in $\gamma \gamma$ collisions i.e. the process $\gamma
\gamma \rightarrow J/\Psi  J/\Psi$ assuming exchange of the 
QCD pomeron (see Fig. 1). It should be noted that both sides 
of the diagram shown in Fig. 1 are characterized by the same (hard) scale 
provided in this case   by the relatively 
large  charmed quark mass.  In this sense this process is complementary to the 
classical measurements listed above.  One of its merits is the fact that 
in this process we can in principle "scan" the perturbative QCD pomeron for arbitrary values of the momentum 
transfer.  In fact the diffractive reaction $\gamma
\gamma \rightarrow J/\Psi  J/\Psi$ is unique in this respect since  
the measurements 
listed above do only probe the QCD pomeron either in forward direction or 
for large momentum transfers.  In the diffractive double $J/\Psi$ production 
in $\gamma \gamma$ collisions the hard scale is provided by the charmed quark 
mass and the theoretical description in terms of the perturbative QCD pomeron 
exchange is applicable for arbitrary momentum transfers.  
  This process has also the advantage that 
its cross-section  can 
be almost entirely calculated perturbatively. The only non-perturbative 
element  is a  parameter determined by the $J/\Psi$ light cone 
wave function  
which can however be obtained from the  measurement of the leptonic 
width $\Gamma _{J/\Psi \rightarrow l^+l^-}$ of the $J/\Psi$.  \\

The imaginary part $Im A(W^2,t=-Q^2)$ of the  amplitude for the process $\gamma
\gamma \rightarrow J/\Psi  J/\Psi$ which corresponds to the
diagram in Fig. 1 illustrating the QCD pomeron exchange can be written in the 
following form: 
\begin{equation}
Im A(W^2,t=-Q^2) = \int {d^2\tdm k\over \pi} {\Phi_0(k^2, Q^2) 
\Phi(x,\tdm k,\tdm Q)\over
[(\tdm k + \tdm Q /2)^2 +s_0][(\tdm k - \tdm Q /2)^2+s_0]} 
\label{ima}
\end{equation}
In this equation $x=m_{J/\Psi}^2/W^2$ where $W$ denotes the total
CM energy of the $\gamma \gamma$ system, $m_{J/\Psi}$ is the
mass of the $J/\Psi$ meson, 
$\tdm Q/ 2 \pm \tdm k$ denote the
transverse momenta 
of the exchanged gluons and $\tdm Q$ is the transverse part of the
momentum transfer.   In the propagators corresponding to the
exchanged gluons we include the parameter $s_0$ which can be
viewed upon as the effective representation of the inverse of
the colour confinement radius squared.  Sensitivity of the
cross-section  to its magnitude can serve as an estimate of the
sensitivity of the 
results to the contribution coming from the infrared region.   
It should be noted that formula 
(\ref{ima}) gives  finite result in the limit $s_0=0$.\\

\noindent
\begin{figure}
\begin{center}
\leavevmode
\epsfxsize = 12cm
\epsfysize = 8cm 
%\epsfbox[5 485 575 795]{ppfig2.ps}
\epsfbox{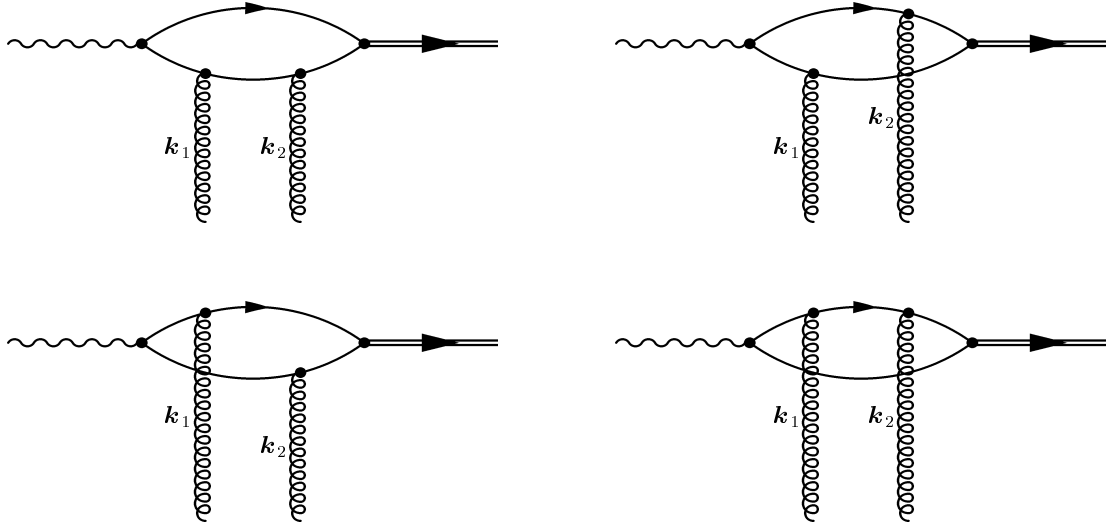}
%\vspace{0.5cm}
\end{center}
\caption{\small
The diagrams describing the coupling of two gluons 
to the $\gamma \rightarrow J/\Psi$ transition vertex. 
}
\end{figure}
 
The impact factor $\Phi_0(k^2, Q^2)$ describes
 the $\gamma J/\Psi$ transition induced by two gluons and the diagrams defining  
 this factor are illustrated in Fig. 2.  
 In the nonrelativistic approximation   they give the following formula for 
 $\Phi_0(k^2, Q^2)$ \cite{FORSHAW,GINZBURG}: 
\begin{equation}
\Phi_0(k^2, Q^2)=
{C\over 2}\sqrt{\alpha_{em}}\alpha_s(\mu^2) \left[{1\over \bar q^2} - 
{1\over m_{J/\Psi}^2/4+k^2}\right] 
\label{impf0}
\end{equation}
where 
\begin{equation}
C=q_c{8\over 3} \pi m_{J/\Psi} f_{J/\Psi}
\label{c}
\end{equation}
with $q_c=2/3$ denoting the charge of a charm quark 
and 
\begin{equation}
\bar q ^2= {m_{J/\Psi}^2+Q^2\over 4}
\label{qbar2}
\end{equation}
The parameter $f_{J/\Psi}$ which characterizes the light cone wave function of
the $J/\Psi$ can be related in the
leading order   to the leptonic width 
$\Gamma_{J/\Psi \rightarrow
l^+l^-}$ of the $J/\Psi$ 

\begin{equation}
f_{J/\Psi}= \sqrt{
{3m_{J/\Psi}\Gamma_{J/\Psi\rightarrow l^+ l^-}
\over 2\pi \alpha_{em} ^2}
} 
\label{fpsi}
\end{equation}

  In our calculations we will set
$f_{J/\Psi}=0.38{\rm \; GeV}$. 
      Equation (\ref{fpsi}) can in principle acquire 
higher order corrections 
 which would affect the normalization factor $C$.  It has however been argued in 
 ref. \cite{FS} that those corrections should be small provided that 
 the light cone wave function  is consistently used.   The apparently large 
 corrections are only present in the non-relativistic potential models where 
 they correspond to the effect of "undressing" the constituent quarks 
  \cite{FS} . \\
  
The function $\Phi(x,\tdm k,\tdm Q)$ satisfies the BFKL equation which in
the leading $ln(1/x)$ approximation has the following form: 
$$    
\Phi(x,\tdm k,\tdm Q)=\Phi_0(k^2, Q^2)+ {3\alpha_s(\mu^2)\over
2\pi^2}\int_x^1{dx^{\prime}\over x^{\prime}} \int
{d^2\tdm k' \over (\tdm k' - \tdm k)^2 + s_0} \times
$$
$$
\left\{\left[{{\tdm k_1^2}\over {\tdm k_1^{\prime 2}} + s_0}   + 
{{\tdm k_2^2}\over {\tdm k_2^{\prime 2}} + s_0} 
  - Q^2 
 {(\tdm k' - \tdm k)^2+s_0 \over ({\tdm k_1^{\prime 2}} + s_0)
 ({\tdm k_2^{\prime 2}} + s_0)}  
\right] 
\Phi(x',\tdm k' ,\tdm Q) - \right.
$$
\begin{equation}
\left. \left[{{\tdm k_1^2}\over {\tdm k_1^{\prime 2}}  + 
(\tdm k' - \tdm k)^2 +2s_0} + 
{{\tdm k_2^2}\over {\tdm k_2^{\prime 2}}  + 
(\tdm k' - \tdm k)^2 +2s_0} \right] 
\Phi(x',\tdm k,\tdm Q) \right\}
\label{bfkl}
\end{equation} 
where 
$$ 
{\tdm k_{1,2}} = {\tdm Q \over 2}\pm \tdm k
$$ 
and 
\begin{equation} 
{\tdm k_{1,2}^{\prime}} = {\tdm Q \over 2} \pm \tdm k^{\prime} 
\label{k12}
\end{equation}
denote the transverse momenta of the gluons.  
The scale of the QCD coupling $\alpha_s$ which appears in
equations (\ref{impf0}) and (\ref{bfkl}) will be  set  
 $\mu^2=k^2+Q^2/4 +m_c^2$ where $m_c$ denotes the mass of the
charmed quark.  The differential cross-section 
is related in the following 
way to the amplitude A:
\begin{equation}
{d \sigma \over dt} = {1\over 16 \pi} |A(W^2,t)|^2
\label{dsdt}
\end{equation}

The BFKL equation (\ref{bfkl}) sums ladder diagrams with
(reggeized) gluon exchange along the ladder. Its kernel contains
therefore the virtual corrections responsible for gluon
reggeization 
 besides the real gluon emission contribution. The former are
given by that part of the integral in the right hand side of the
equation (\ref{bfkl}) whose integrand is proportional to 
$\Phi(x',\tdm k, \tdm Q)$ while the latter corresponds to the
remaining part of the integral.  If in eq. (\ref{ima}) one approximates 
the function 
$\Phi(x,\tdm k,\tdm Q)$ by  the impact factor $\Phi_0(k^2, Q^2)$ 
then one gets the two (elementary) gluon exchange contribution to the process
$\gamma \gamma \rightarrow J/\Psi J/\Psi$ \cite{GINZBURG}.  
The two gluon exchange
mechanism gives the cross-section  which is independent of
energy. A possible increase of the cross-section with energy is
  described by the BFKL effects generated by the solution
of  equation (\ref{bfkl}).  These effects can also
significantly affect the $t$ dependence of the cross-section   
and so the process $\gamma \gamma \rightarrow 
J/\Psi J/\Psi$ might be a useful tool for probing the
t-dependence  which follows from the BFKL equation. 
Let us recall that in the diffractive photo-production of
$J/\Psi$ on a proton a possible nontrivial $t$-dependence generated
by the BFKL equation cannot be detected due to (non-perturbative)
coupling of the two gluon system to a proton \cite{JPSIP}.  
In order to avoid this effect one may consider the diffractive vector meson 
photoproduction on a proton accompanied 
by proton dissociation \cite{FORSHAW,BARTLQ}.  In this case however 
the momentum transfer has to be large. \\

It is known that the BFKL equation can acquire significant
non-leading contributions \cite{BFKLNL,DGROSS,SALAM}.  Although the structure
of those 
corrections is fairly complicated their dominant part is
rather simple and follows from restricting the integration
region in
the real 
emission term in  
equation (\ref{bfkl}).  For $Q=0$ the relevant limitation is 
\cite{KMSTAS,KMSG,KMSG1} 
\begin{equation}
k'^2 \le k^2 {x'\over x} 
\label{kc1}
\end{equation}   
It follows from  the requirement that the virtuality of the
gluons exchanged along the chain is dominated by the transverse
momentum squared.  
  The  constraint (\ref{kc1}) can be shown to
exhaust about $70 \%$ of the next-to-leading corrections 
to the QCD pomeron intercept \cite{BFKLNL,KMSG1}.  Generalization of the
constraint (\ref{kc1}) to the case of a non-forward configuration
with $Q^2 \ge 0$ takes the following form: 
\begin{equation} 
k'^2 \le (k^2+ Q^2/4) {x'\over x}
\label{kc2}
\end{equation} 
Besides the BFKL equation (\ref{bfkl}) in the leading
logarithmic approximation we shall therefore also consider the 
equation which will embody the constraint
(\ref{kc2}) in order to estimate possible effect of the non-leading
contributions. \\

The corresponding equation which contains constraint (\ref{kc2}) in the
real emission term reads: 
$$    
\Phi(x,\tdm k,\tdm Q)=\Phi_0(k^2, Q^2)+ {3\alpha_s(\mu^2)\over
2\pi^2}\int_x^1{dx^{\prime}\over x^{\prime}} \int
{d^2\tdm k' \over (\tdm k' - \tdm k)^2 + s_0} \times
$$
$$
\left\{\left[{{\tdm k_1^2}\over {\tdm k_1^{\prime 2}} + s_0}   + 
{{\tdm k_2^2}\over {\tdm k_2^{\prime 2}} + s_0} 
  - Q^2 
 {(\tdm k' - \tdm k)^2+s_0 \over ({\tdm k_1^{\prime 2}} + s_0)
 ({\tdm k_2^{\prime 2}} + s_0)}  
\right] \times \right.
$$ 
$$ 
\Phi(x',\tdm k' ,\tdm Q) \Theta \left((k^2+Q^2/4)x'/x-k^{\prime 2}) \right) 
 - 
$$
\begin{equation}
\left. \left[{{\tdm k_1^2}\over {\tdm k_1^{\prime 2}}  + 
(\tdm k' - \tdm k)^2 +2s_0} + 
{{\tdm k_2^2}\over {\tdm k_2^{\prime 2}}  + 
(\tdm k' - \tdm k)^2 +2s_0} \right] 
\Phi(x',\tdm k,\tdm Q) \right\}
\label{bfklkc}
\end{equation}

We solved  equations (\ref{bfkl}) and (\ref{bfklkc}) numerically setting 
$m_c=m_{J/\Psi} /2$, $\Lambda_{QCD}=0.23 {\rm \; GeV}$ and using the one 
loop approximation for the QCD coupling $\alpha_s$ with the number of flavours
$N_f=4$.   Brief summary of the 
numerical method and of the adopted approximations in solving 
equations (\ref{bfkl},\ref{bfklkc})  will be given below.   
Let us recall that we used a running coupling with the scale 
$\mu^2=k^2+Q^2/4+m_c^2$. The parameter $s_0$ was varied within the range 
$0.04 {\rm \; GeV}^2 < s_0<0.16 {\rm \; GeV}^2$.  It should be noted 
that  the solutions of 
equations (\ref{bfkl}, \ref{bfklkc}) and the amplitude (\ref{ima}) are 
finite in the limit $s_0=0$. This follows from the fact that both impact 
factors $\Phi_0(k^2, Q^2)$ and $\Phi(x,\tdm k,\tdm Q)$ vanish for 
$\tdm k=\pm \tdm Q/2$ (see equations (\ref{impf0}, \ref{bfkl}, \ref{bfklkc})).
The results with finite $s_0$ are however  more realistic. \\

\noindent
\begin{figure}[hbpt]
\begin{center}
\epsfxsize = 13cm
\epsfysize = 13cm 
\leavevmode
\epsfbox[18 200 565 755]{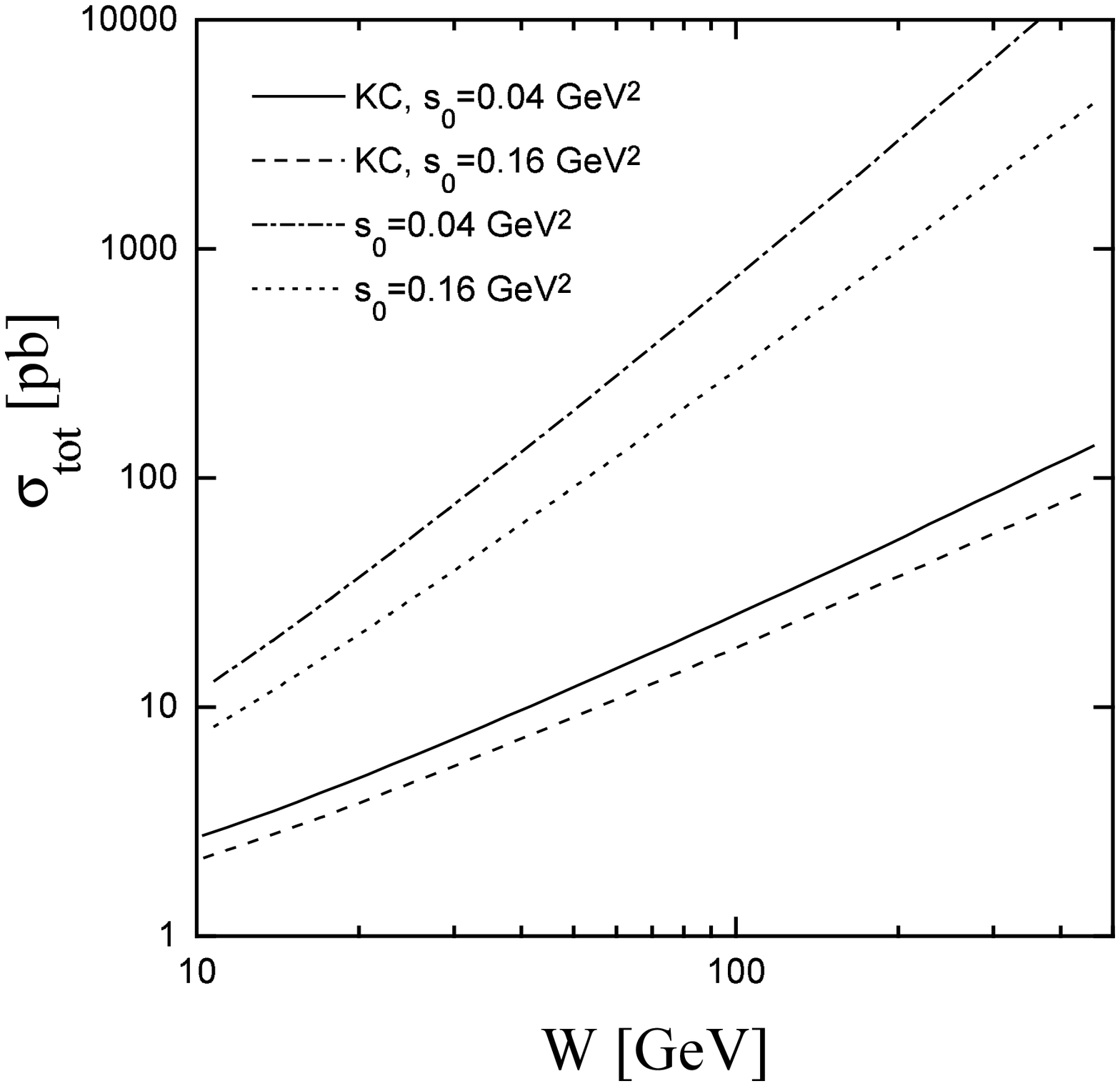}
\end{center}
\caption{\small
Energy dependence of the cross-section for the process 
$\gamma\gamma \rightarrow J/\Psi J/\Psi$.  The two lower curves
correspond to the calculations based on  equation
(\ref{bfklkc0}) which contains the non-leading effects coming
from the  constraint (\ref{kc2}).  The continuous line
corresponds to $s_0=0.04 {\rm \; GeV}^2$ and the dashed line to $s_0=0.16
{\rm \; GeV}^2$.  The two upper curves correspond to equation
(\ref{bfkl}) i.e. 
to the BFKL equation in the leading logarithmic approximation.  
The dashed-dotted line corresponds to $s_0=0.04 {\rm \; GeV}^2$ 
and short dashed line to $s_0=0.16{\rm \; GeV}^2$. 
}
\end{figure}

For fixed (i.e. non-running) coupling and for $s_0=0$  equations 
(\ref{bfkl},\ref{bfklkc}) could in principle be solved analytically taking advantage of  
the conformal invariance of their kernels.    
The numerical method which we adopted is however more flexible and allows 
to analyze equations (\ref{bfkl},\ref{bfklkc}) in the more realistic case 
of running $\alpha_s$ and non-zero $s_0$.\\

  In order to solve equations 
 (\ref{bfkl},\ref{bfklkc}) we at first expand the function 
$\Phi(x,\tdm k,\tdm Q)$  in the (truncated) Fourier series 
\begin{equation}
\Phi(x,\tdm k,\tdm Q)=\sum_0^N \tilde \Phi_m(x, k^2,Q^2) cos(2m \phi)
\label{fourier}
\end{equation}
where $\phi$ denotes the azimuthal angle between the (two-dimensional) vectors $\tdm k$ 
and $\tdm Q$ and then we discretize the corresponding system of integral 
equations for the functions $ \tilde \Phi_m(x, k^2,Q^2)$   
by the Tchebyshev interpolation method.  We found that  the 
BFKL equation  (\ref{bfkl})  can be to a very good accuracy approximated 
by retaining only the term corresponding to $m=0$ that corresponds to neglecting 
possible dependence of the function  $\Phi(x,\tdm k,\tdm Q)$ upon 
the azimuthal angle $\phi$.  
After retaining only the term $ \tilde \Phi_0(x, k^2,Q^2)$  one gets the following equation  for this function:  
$$
 \tilde \Phi_0(x, k^2,Q^2)=\Phi_0(k^2,Q^2) + 
$$
$$
 {3 \alpha_s(\mu^2) \over \pi} 
 \int_x ^1 {dx'\over x'} \left[ \int_0^{\infty} dk'^2 R(k^2,k^{\prime 2},Q^2) 
 \tilde \Phi_0(x', k'^2,Q^2)-\right.
$$
\begin{equation}  
 \left. V(k^2,Q^2)\tilde \Phi_0(x', k^2, Q^2)\right] 
\label{bfkl0}
\end{equation}
where 
$$
R(k^2,k^{\prime 2},Q^2)={1\over 
\sqrt{(Q^2/4+ k'^2 +s_0)^2 - Q^2 k'^2}} \left \{ 
{1 \over \sqrt{(k^2+ k'^2 +s_0)^2 - 4 k^2 k'^2}} \right.
$$
$$
 \left[(k^2 +Q^2/4)-{Q^2 k^2\over
 \sqrt{(Q^2/4+k'^2 +s_0)^2- Q^2 k'^2}+
  Q^2/4+k'^2 +s_0}\right] 
$$
\begin{equation}
\left. -{Q^2\over (Q^2/4+k'^2 +s_0)}\right\}
\label{real}
\end{equation}
and 
$$
V(k^2,Q^2)=\int_0^{\infty} dk^{\prime 2} {1\over  
\sqrt{(Q^2/4+ k'^2 +s_0)^2 - Q^2 k'^2}} \left \{ 
{1 \over \sqrt{(k^2+ k'^2 +s_0)^2 - 4 k^2 k'^2}} \right.
$$
$$
 \left. \left[(k^2 +Q^2/4)-{Q^2 k^2\over
 \sqrt{(Q^2/4+k'^2 +s_0)^2- Q^2 k'^2}+
  Q^2/4+k'^2 +s_0}\right] \right\}+ 
$$  
\begin{equation} 
\int_0^{2\pi} {d \phi\over 2 \pi} \int_0^1 d\lambda 
 {k^2 + Q^2/4 +kQ cos(\phi) \over 
 [k^2 +Q^2/4 + kQcos(\phi) +(1+\lambda)^2 s_0]} \vspace{1em}
\label{virt}
\end{equation}
We assume that similar approximation can be adopted in the solution 
of eq. ({\ref{bfklkc}).     
Equation (\ref{bfklkc}) then takes a 
similar form to equation (\ref{bfkl0}) with additional constraint 
$\Theta[(k^2+Q^2/4) x'/x-k'^2]$ imposed on the real emission terms i.e. 
on those terms in eq. (\ref{bfkl0}) in which the corresponding
integrands  
contain the factors $\tilde \Phi_0(x',k'^2, Q^2)$.  The corresponding 
equation  reads: 
$$
 \tilde \Phi_0(x, k^2,Q^2)=\Phi_0(k^2,Q^2) + 
$$
$$
 {3 \alpha_s(\mu^2) \over \pi} 
 \int_x ^1 {dx'\over x'} \left[ \int_0^{\infty} dk'^2 R(k^2,k^{\prime 2},Q^2) 
 \tilde \Phi_0(x', k'^2,Q^2)\Theta \left((k^2+Q^2/4) x'/x-k'^2\right)-\right. 
$$ 
\begin{equation}   
\left. V(k^2,Q^2)\tilde \Phi_0(x', k^2, Q^2)\right] 
\label{bfklkc0}
\end{equation}

We based our calculations on the solutions of  equations
(\ref{bfkl0}, \ref{bfklkc0}).   
In Fig. 3 we show the cross-section for the process 
$\gamma \gamma \rightarrow J/\Psi J/\Psi$ 
plotted as  function of the total CM energy $W$.   
We show results based on the BFKL equation in the leading logarithmic 
approximation as well as those which include the dominant non-leading 
effects.  The calculations were performed for the two values of the parameter 
$s_0$ i.e. $s_0=0.04 {\rm \; GeV}^2$ and $s_0=0.16 {\rm \; GeV}^2$. 
We have also estimated the total cross-section for the process 
$e^+e^-\rightarrow e^+e^-  J/\Psi J/\Psi$ with antitagged $e^+$ and $e^-$   
 for
the LEP2 energies assuming
the same cuts   as in ref. \cite{GGREV}.  We get  
$\sigma_{e^+e^-\rightarrow e^+e^-  J/\Psi J/\Psi}(\sqrt{s}=175
GeV)=0.12$ pb and 
$\sigma_{e^+e^-\rightarrow e^+e^- 
J/\Psi J/\Psi}(\sqrt{s}=175)=0.09$ pb 
for $s_0 =$ 0.04 GeV$^2$ and $s_0 =$ 0.16 GeV$^2$ respectively.\\   

In Fig.4 we show the $t$-dependence of the cross-section calculated for 
$s_0 = 0.10 {\rm \; GeV}^2$.
We show in this Figure results for two values of the CM energy $W$ 
($W=50 {\rm \; GeV}$  and $W=125 {\rm \; GeV}$) 
obtained from the solution of the BFKL equation with the non-leading effects taken into 
account (see eq. (\ref{bfklkc0})) and confront them with 
the Born term which corresponds to the two (elementary) gluon exchange.  
The latter is of course independent of the energy $W$.  
The values of the energy $W$ were chosen to be in the region which may 
be accessible at LEP2. The following points should be emphasized: 
\begin{enumerate} 
\item We see from Fig. 3 that the effect of the non-leading contributions is 
very important and that they  significantly reduce  magnitude of the 
cross-section and slow down its increase with increasing CM
energy $W$.    

\item 
The magnitude of the cross-section decreases with increasing
magnitude of the  parameter $s_0$ which controls the
contribution coming from the infrared region.  This effect is
however much weaker than that generated by the  
constraint (\ref{kc2}) which gives the dominant non-leading
contribution.  The  energy dependence of the
cross-section is  practically unaffected  by the parameter $s_0$.

\item 
It can be seen from Fig. 3 that the cross-section exhibits an  
approximate $(W^2)^{2\lambda}$ dependence. The parameter $\lambda$ 
slowly increases  with increasing energy $W$ and changes from
$\lambda \approx 0.23$ at $W=20$ GeV to $\lambda \approx 0.28$ at 
$W=500$ GeV i.e. within the energy range which is relevant for
LEP2 
and for possible TESLA measurements.
  These results correspond to the solution of
the BFKL equation (\ref{bfklkc0}) which contains the non-leading
effects generated by the constraint (\ref{kc2}).      
The (predicted) energy dependence of the cross-section 
 ($(W^2)^{2\lambda}, \lambda \sim 0.23 - 0.28$)  
is  marginally steeper than  that observed in 
$J/\Psi$ photo-production \cite{VMPHOTOP}. It should however be remembered 
that the non-leading 
effects which we have taken into account although being  the dominant ones 
still do not 
exhaust all next-to-leading QCD corrections to the BFKL kernel \cite{BFKLNL}.  
The remaining contributions are expected to reduce the parameter $\lambda$ 
but their effect may be expected to be less important than that generated 
by the constraint (\ref{kc2}).  
 The   BFKL equation in the leading 
logarithmic approximation generates a much stronger energy dependence
of the cross-section (see Fig. 3). 
 
\item 
The enhancement of the cross-section  is still appreciable  after 
including 
the dominant non-leading contribution which follows from the
constraint (\ref{kc2}). Thus while in the Born approximation 
(i.e. for the elementary two gluon exchange 
which gives an energy independent cross-section)  
 we get $\sigma_{tot} \sim 1.9-2.6$~pb the cross-section calculated from the 
 solution of the BFKL equation with the non-leading effects taken into account 
can reach the value 4~pb at $W=20$ GeV and 26~pb for $W=100$ GeV i.e. for 
energies which can be accessible at LEP2.

\item 
Plots shown in Fig.~4 show that the BFKL effects significantly affect 
the $t$-dependence of the differential cross-section leading to steeper 
$t$-dependence than that generated by the Born term.   Possible
energy dependence of the diffractive slope is found to be very
weak (see Fig. 4). A similar result was also found in the BFKL equation 
in the leading logarithmic approximation \cite{BARTLQ}. 

\item
 The variation of the parameter $s_0$ within
the range $0.04$ GeV$^2<s_0<0.16$ GeV$^2$ changes normalization of the
cross-section by about 30 \%.  There may still be other sources of   
normalization uncertainties coming for instance from  the use of 
the nonrelativistic approximation of the impact
factor etc. which can increase the normalization error up to 50\% or so.
 The energy dependence of the cross-section is however an
unambigous theoretical prediction.    
\end{enumerate}

\begin{figure}
\begin{center}
\epsfxsize = 13cm
\epsfysize = 13cm 
%\epsfbox[18 200 565 75]{ppfig4.ps}\\
\leavevmode
\epsfbox{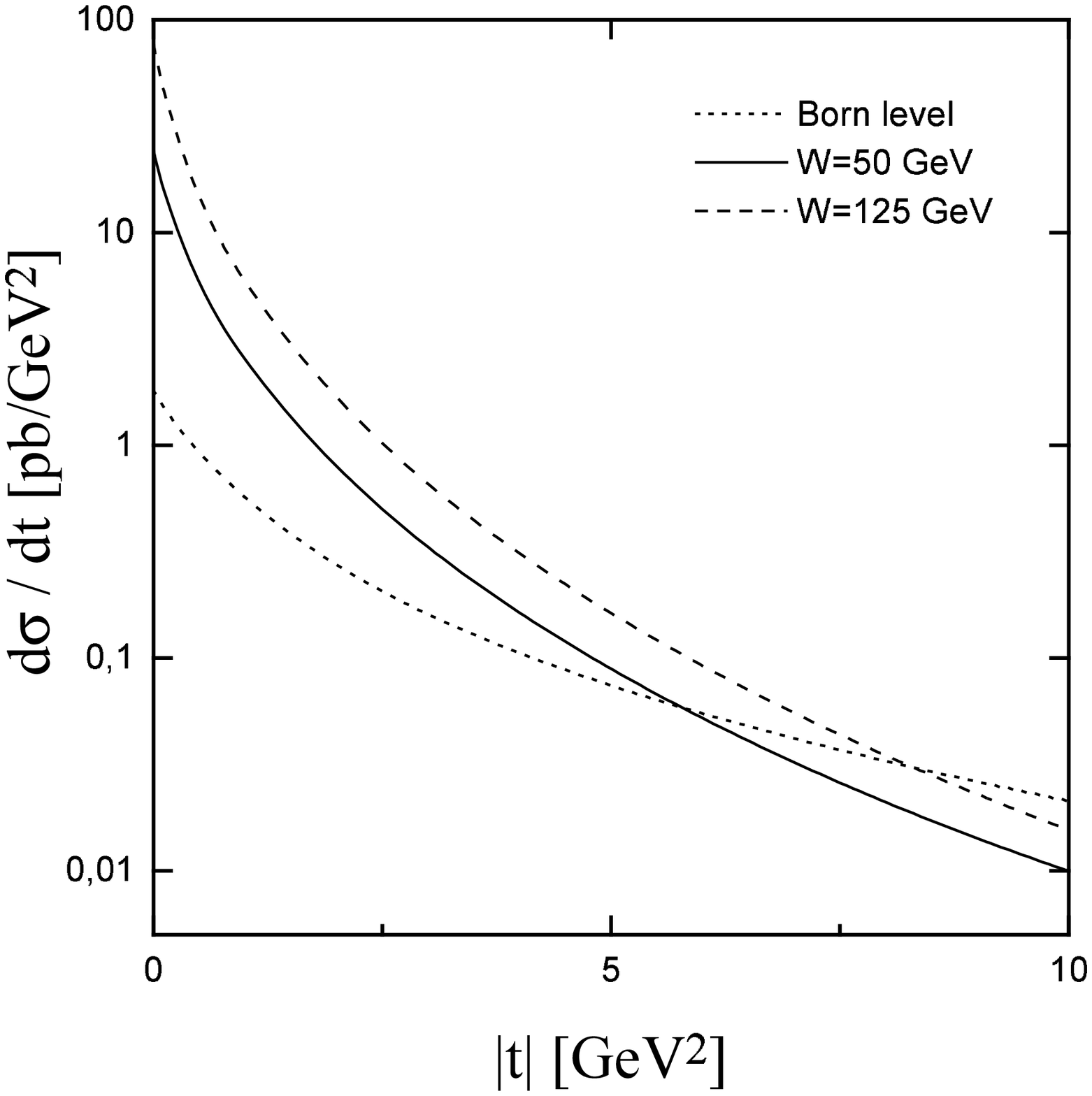}
\end{center}
\caption{\small
The differential cross-section of the process $\gamma
\gamma \rightarrow J/\Psi J/\Psi$ corresponding to the solution 
of  equation (\ref{bfklkc0}) which contains the non-leading effects coming
from the kinematical constraint (\ref{kc2}) shown for two 
values of the CM energy $W$, $W=50{\rm \; GeV}$ (continuous line) and 
$W=125 {\rm \; GeV}$ (dashed line).  The short dashed line corresponds 
to the Born term i.e. to the elementary two gluon exchange
mechanism which gives the energy independent cross-section. 
The parameter $s_0$ was set equal to $0.10{\rm \; GeV}^2$. 
}
\end{figure}

In our calculations we have assumed dominance of the imaginary part of the 
production amplitude.  The effect of the real part can be taken into account by
multiplying the cross-section by the correction factor $1+tg^2(\pi\lambda/2)$ 
which for $\lambda \sim 0.25$ can introduce additional enhancement of 
about 20~\%.\\

It may finally be instructive to confront our results with
recent findings concerning the solution of the BFKL equation in
the next-to-leading approximation \cite{BFKLNL,DGROSS,SALAM}.  
It has been found that 
the non leading effects are very important and that the effective intercept 
$\lambda$ can become negative in the next-to-leading approximation 
for  relevant values of $\alpha_s > 0.15$.  It has been argued that 
the next-to-leading approximation is not reliable and that one has to perform 
complete resummation of the non-leading contributions \cite{DGROSS,SALAM}.   
Let us observe that equation (\ref{bfklkc0}) resums to all orders non-leading 
effects generated by the
contraint (\ref{kc2}).  The difference between exact 
solution of this equation and its next-to-leading approximation was 
discussed in ref. \cite{KMSG1} for $t=0$ and for the fixed coupling
$\alpha_s$.  It has in particular been found  that the exponent
$\lambda$ 
corresponding to the exact solution stays always positive for
arbitrary values of the coupling $\alpha_s$.  The next-to-leading approximation 
for this exponent differs significantly from the exact solution already 
for $\alpha_s > 0.2$ and can again become negative.  This result confirms 
the observation \cite{DGROSS,SALAM} that the next-to-leading approximation 
alone is unreliable and that one has to perform complete resummation of the 
non-leading effects.\\

To sum up we have developed the formalism that enabled us to estimate the 
contribution of the QCD pomeron in the 
$\gamma \gamma \rightarrow J/\Psi J/\Psi$ diffractive production process.  
We found that the BFKL effects give a significant enhancement of the 
cross-section and modify the $t$-dependence of the Born term.  
The cross-section exhibits an approximate power law dependence  
$(W^2)^{2\lambda}$ with $\lambda \sim 0.25$.  
We based our calculations on the BFKL equation which contained  non-leading 
contributions coming from the constraint 
imposed upon the available phase space.  This constraint is the dominant 
non-leading effect and it exhausts  about    
$70 \%$ of the next-to-leading corrections to the BFKL pomeron intercept. 
We found that the non-leading contributions generated by constraint 
(\ref{kc2})  significantly affect theoretical 
expectations based on the BFKL equation in the leading logarithmic
approximation. This means that the enhancement of the cross-section although 
still quite appreciable should be much smaller than that which follows from 
estimates based on the leading logarithmic approximation \cite{GGREV}. 

\section*{Acknowledgments}
We thank Albert De Roeck for his interest in this work and useful 
discussions. L.M. is grateful to the Foundation for Polish Science for a
fellowship. This research was partially supported
by the Polish State Committee for Scientific Research (KBN) grants
2~P03B~184~10, 2~P03B~89~13, 2~P03B~044~14, 2~P03B~084~14 and by the 
EU Fourth Framework Programme 'Training and Mobility of Researchers', Network 
'Quantum Chromodynamics and the Deep Structure of Elementary Particles', contract FMRX - CT98 - 0194.

\end{document}